# On the internalisation, intraplasmodial carriage and excretion of metallic nanoparticles in the slime mould *Physarum polycephalum*

Richard Mayne[1,2], David Patton[1,] Ben de Lacy Costello[1,2], Andrew Adamatzky[2], Rosemary Camilla Patton[1]

[1] Faculty of Health and Life Sciences, University of the West of England, Bristol, UK
[2] Unconventional Computing Centre, University of the West of England, Bristol, UK

**ABSTRACT**

The plasmodium of *Physarum polycephalum* is a large single cell visible with the naked eye. When inoculated on a substrate with attractants and repellents the plasmodium develops optimal networks of protoplasmic tubes which span sites of attractants (i.e. nutrients) yet avoid domains with a high nutrient concentration. It should therefore be possible to program the plasmodium towards deterministic adaptive transformation of internalised nano- and micro-scale materials. In laboratory experiments with magnetite nanoparticles and glass micro-spheres coated with silver metal we demonstrate that the plasmodium of *P. polycephalum* can propagate the nano-scale objects using a number of distinct mechanisms including endocytosis, transcytosis and dragging. The results of our experiments could be used in the development of novel techniques targeted towards the growth of metallised biological wires and hybrid nano- and micro-circuits.

**INTRODUCTION**

The plasmodium of *Physarum polycephalum* (Order *Physarales*, class *Myxomecetes*, subclass *Myxogastromycetidae*) is a single macroscopic cell which can grow to tens of centimetres in length (Stephenson and Stempen, 1994). The plasmodium feeds on bacteria and microscopic food particles by endocytosis; when placed in an environment with distributed sources of nutrients, the plasmodium migrates between each source via contraction of muscle proteins (Stephenson and Stempen, 1994), forming a network of interconnected protoplasmic tubes. Nakagaki *et al.* (2001) demonstrated that the topology of the plasmodium's protoplasmic network optimizes resource harvesting and increases the efficiency of intraplasmodial transport from the nutrient source to the rest of the plasmodium. In Adamatzky (2010a), we have shown how to construct specialised and general purpose massively-parallel amorphous computers from the plasmodium of *P. polycephalum* that are capable of solving problems of computational geometry, graph-theory and logic.

When the slime mould develops a network of protoplasmic tubes spanning sources of nutrients, the cell maintains its integrity by pumping nutrients and metabolites between remote parts of its body via cytoplasmic streaming (Allen *et al.*, 1963; Bykov *et al.*, 2009; Gawlitta *et al.*, 1980; Hulsmann and Wohlfarth-Bottermann, 1978; Newton *et al.*, 1977; Stewart and Stewart, 1959). This cytoplasmic streaming may be manipulated experimentally and employed for the transportation of exogenous bio-compatible substances inside the protoplasmic network: in Adamatzky (2010b) we demonstrated that the plasmodium of *P. polycephalum* consumes various coloured dyes and distributes them throughout its protoplasmic network. By specifically arranging a configuration of attractive (sources of nutrients) and repelling (sodium chloride crystals) fields, it is possible to program the plasmodium to implement the following operations: to take in specific coloured dyes from the closest coloured oat flake; to mix two different colours to produce a third colour; and, to transport colour to a specified locus of an experimental substrate. Transportation of colourings *per se* is of little interest but shows the potential of *P. polycephalum* as a programmable transport medium.

We propose that *P. polycephalum's* potential for internalisation and re-distribution of foreign particles may be employed for the development of self-growing electronic circuits and/or hybrid slime mould-artificial computing devices, providing that the plasmodium may be manipulated to internalise and propagate suitable artificial circuit components. Inspired by our previous results (Adamatzky, 2010b) and studies on cellular endocytosis of magnetic nano-beads (Li et al., 2005) and fluorescent nano-beads (Bandmann et al., 2012), nanowire scaffolding for living tissue (Tian et al., 2012) and intake of latex beads by the amoeba's endocytotic mechanisms (Goodall and Thompson, 1971), we chose to focus on two types of materials: magnetic magnetite nanoparticles - because they can be used for *in situ* construction of basic nano-scale electronic devices - and glass microspheres coated with silver, because they have be used to grow conductive pathways, i.e. hybrid bio-artificial wires.

**METHODS**

**Culture and Nanoparticle Intake**

Plasmodium of *P. polycephalum* was cultivated in plastic containers, on paper towels sprinkled with distilled water and fed with oat flakes (Asda's Smart Price Porridge Oats, UK). *P. polycephalum* was cultured experimentally upon 2% non-nutrient agar gel (Select agar, Sigma Aldrich, UK) poured in 9 cm plastic Petri dishes. In each experiment, an oat flake colonised by plasmodium was placed in the centre of the Petri dish and 200μl of nanoparticle suspension was dispensed onto the colony. A circle with a radius of 1-1.5cm was cut around the inoculation site to prevent the suspension diffusing into the surrounding gel. The following two types of nanoparticles were used:
1. ChemiCell FluidMAG-D superparamagnetic magnetite (Iron II/III oxide) nanoparticles (ChemiCell GmbH, Berlin, Germany), 200nm diameter. The particles were coated with a hydrodynamic with starch matrix (Fig. 1) to protect them against aggregation by foreign ions (Chemicell GmbH, 2012).
2. Silver coated hollow glass microspheres, 10-27μm diameter, suspended in Tween 80 biocompatible surfactant (Cospheric, Santa Barbara, USA) (Fig. 2).

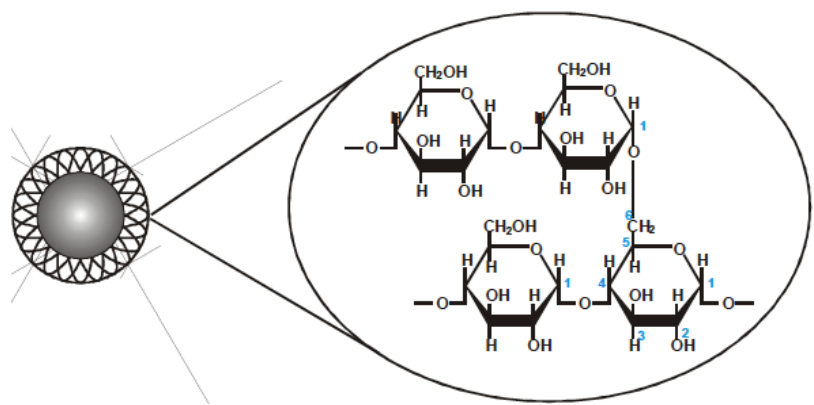

**Fig. 1.** Magnetite nano-particles in starch matrix. From [fluidMAG-D, 2012].

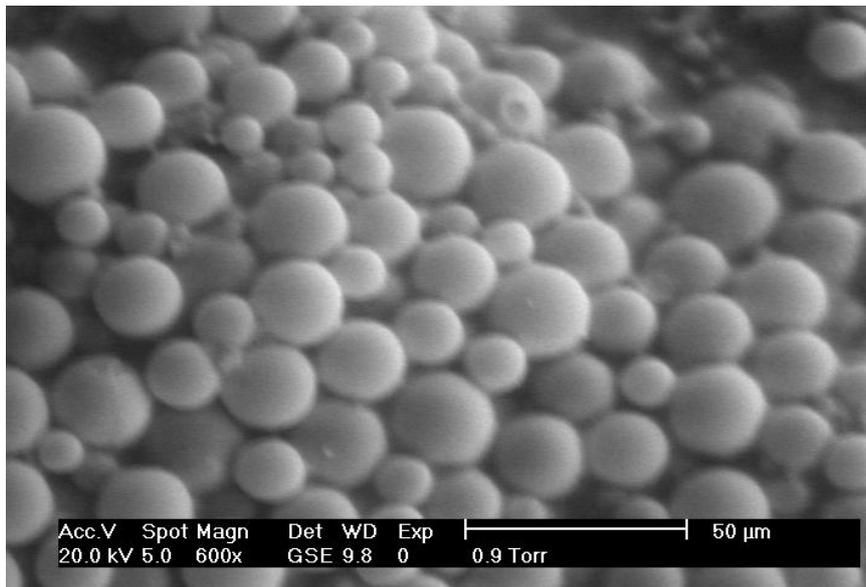

**Fig. 2.** Scanning electron micrograph showing silver coated hollow glass spheres 10-27 μm diameter in the loading zone.

To facilitate directional growth of a plasmodial network, oat flakes were placed near the perimeter of the Petri dish. Protoplasmic networks were sampled 3-4 cm away from the inoculation site and analysed with scanning electron microscopy (SEM), energy-dispersive x-ray microanalysis (EDX) and transmission electron microscopy (TEM) studies.

**Scanning Electron Microscopy and Energy-Dispersive X-ray Microanalysis**

Whole plasmodia were air-dried before samples of plasmodial tubes (thin, interlinking projections of plasmodium which link food sources) were dissected out of their plates and mounted on SEM stubs using double sided conductive carbon tabs. They were then sputter coated with gold before being viewed in a Philips XL30 environmental scanning electron microscope (ESEM). Fully hydrated plasmodium was also observed by placing pieces of agar on a Peltier cooling stage at 5$^{o}$C and observing at pressures between 4.5 and 6.5 Torr.

Semi-thin resin sections of fixed plasmodial tubes (see next section) were mounted upon stubs for SEM imaging and EDX using an Oxford Instruments Link system to identify elements.

**Transmission Electron Microscopy**

Plasmodium samples were fixed in 2% glutaraldehyde in pH 7.1 potassium phosphate buffer for 1 hr at room temperature. They were rinsed x3 for 1 hr in the same buffer in a fridge. They were postfixed in 1% osmium tetroxide (aq.) in the same buffer for 1 hr. Samples were embedded in TAAB embedding Resin (TAAB, Aldermaston, UK). Sections were cut using a Reichert-Jung Ultracut E ultramicrotome. Semi-thin sections were stained with toluidine blue. Ultra-thin sections were stained with uranyl acetate and lead citrate before being viewed in a Philips CM10 TEM.

Images of dried nanoparticles and microspheres were produced by diluting suspensions with deionised water and dispensing them on to copper specimen grids, before drying in an oven at 60$^{o}$C.

**RESULTS**

**Magnetite**

Magnetic nanoparticles were internalised by the plasmodium of *P. polycephalum*. Slime mould treated with nanoparticle suspension was noted to behave slightly differently to controls; they did not always grow towards oat flakes and assumed a more diminutive morphology (Fig. 3).

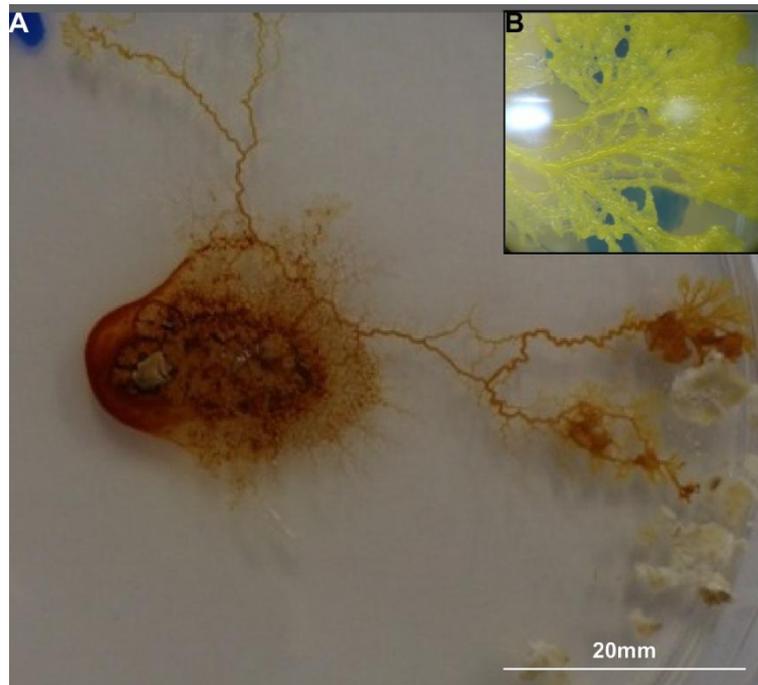

**Fig. 3.** Images of the macroscopic appearance of slime mould. (A) Slime mould loaded with fluidMAG-D magnetic nanoparticles. (B) Control slime mould.

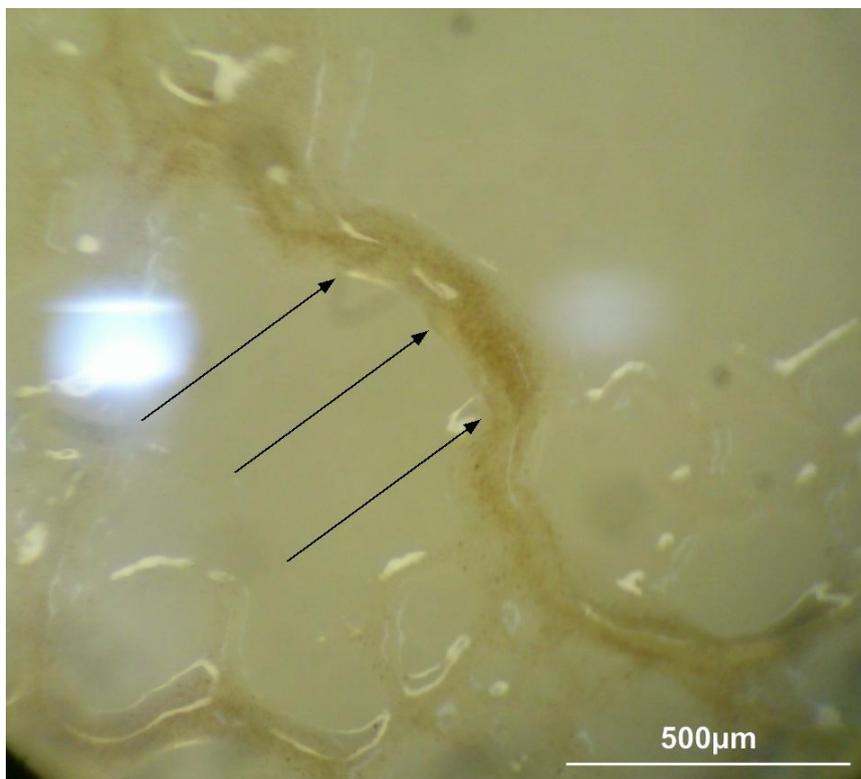

**Fig 4**. Dissecting light micrograph showing surface detail of unfixed protoplasmic network; aggregates of magnetite particles are clearly visible in the tubes. Major particle filled tube is highlighted by arrows.

Plasmodia exposed to magnetite nanoparticles were densely stained (Fig. 3) and appeared to contain particulate material when viewed with a dissecting light microscope (Fig. 4). Light microscopic observation of semi-thin transverse sections of plasmodium revealed plentiful, irregularly dispersed, discrete dark orange polygons (Fig. 5). These foci were also visible using SEM, and were revealed to have a high atomic number content, as denoted by their bright appearance when viewed using a back scattered electron detector (Fig. 6). These regions were consequently found to contain high levels of iron using EDX (Fig. 7), suggesting that they were indeed accumulations of magnetite nanoparticles. SEM imaging of dried tube segments revealed small, irregularly dispersed foci of EDX-confirmed iron-containing deposits on the surface of the tube (data not shown).

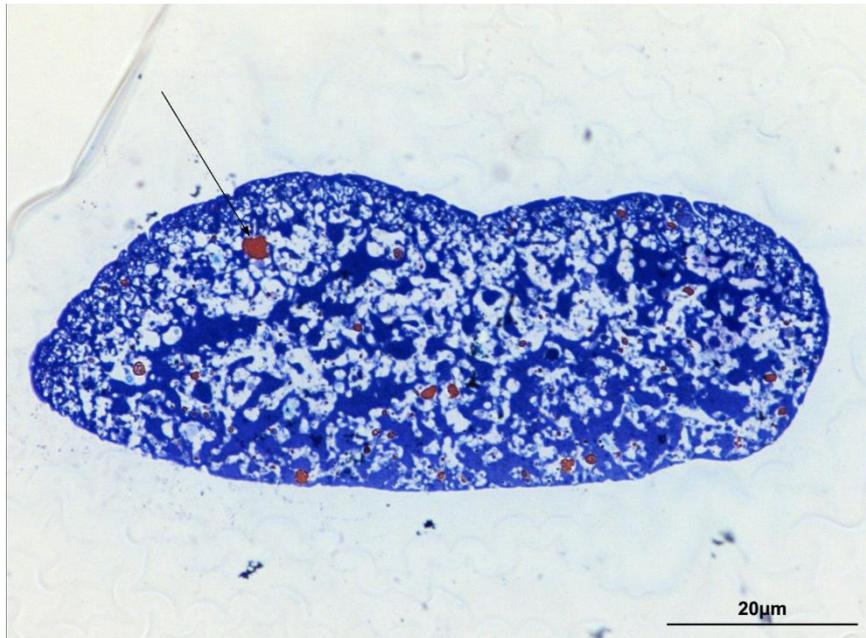

**Fig. 5.** Cross-section of protoplasmic tube with aggregates of magnetic 200 nm particles visible as dark-grey polygons, the largest of which is highlighted by an arrow. Toluidine blue stain.

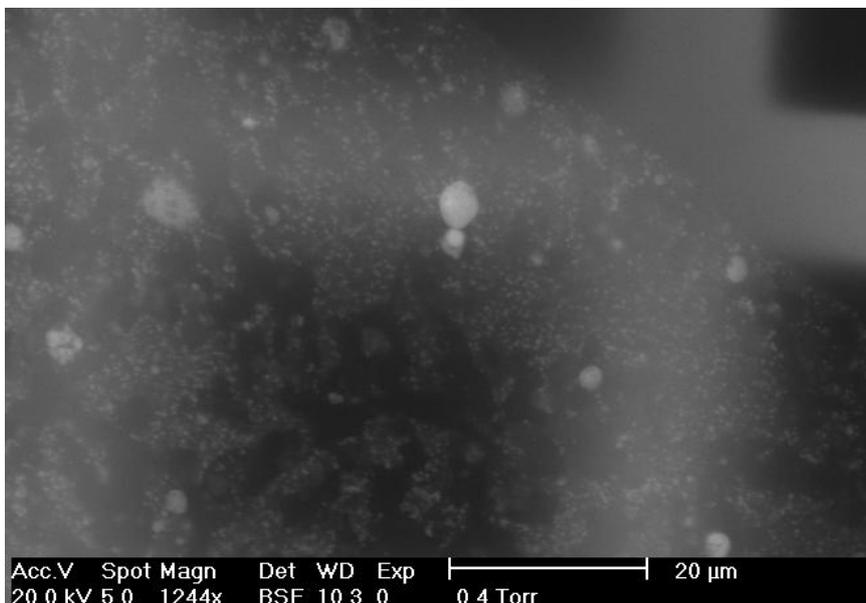

**Fig. 6.** SEM micrograph of a semithin section of protoplasmic tube from plasmodium treated with nanoparticle suspension; bright spots correspond to areas of high atomic number content.

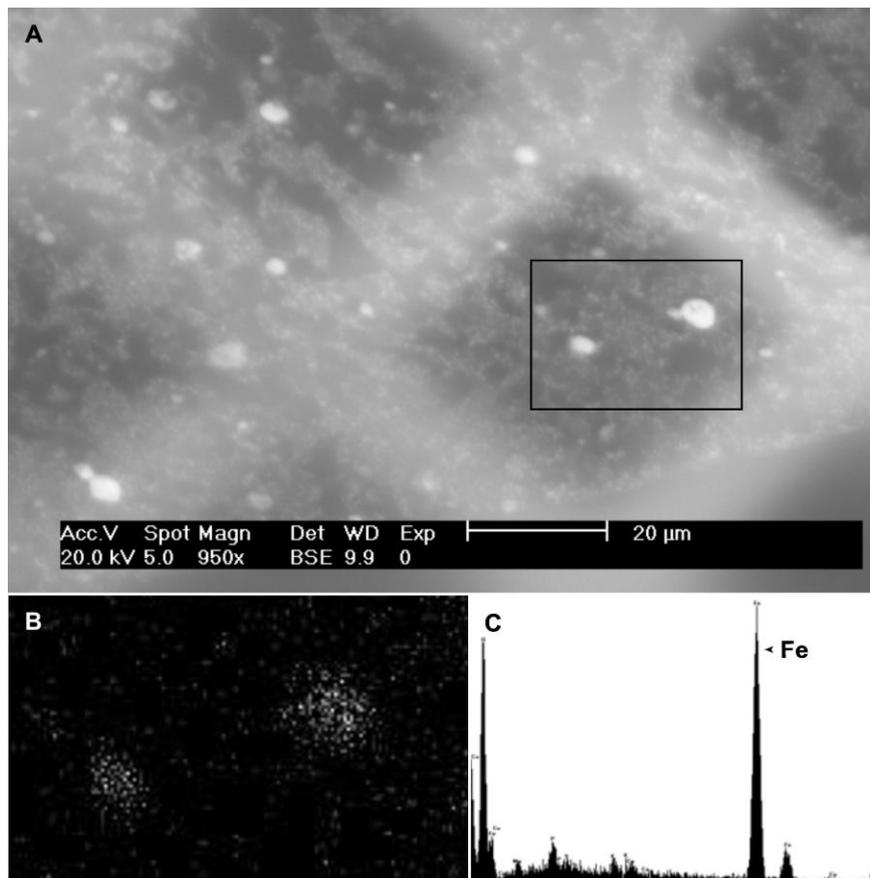

**Fig. 7.** Scanning electron micrograph of plasmodial tube with corresponding EDX map and spectrum, demonstrating high amounts of iron in bright regions (A) SEM image; area denoted by box corresponds to EDX map in (B). (B) EDX map showing iron x-ray emission corresponding with bright areas in (A). (C) EDX spectrum of bright region, showing a significant iron peak.

Dried nanoparticle suspensions tended to aggregate. When observed using TEM, possible 200nm sized regions representing the multicore particles could be seen at the edges of samples (Fig. 8 A). These regions are seen to be composed of smaller nanoparticles in more diluted samples (Fig. 8 B). When inside slime mould plasmodia, magnetite aggregates of 1- 5µm were unevenly distributed throughout the whole plasmodium. These aggregates had a tendency to be contained within vacuoles when in the peripheral region of the tube (the ectoplasm), wherein they were also more frequent and larger than deposits in the centre (the endoplasm) (Fig. 9 A). Conversely, deposits in the endoplasm were frequently continuous with the protoplasm (Fig. 9 B). Beyond the magnetite aggregates, small 5-20nm particles could be seen in the endoplasm in higher magnification images (Fig 9 B); the appearance of these objects was highly similar to the individual components of the multicore particles seen in Figure 8 B.

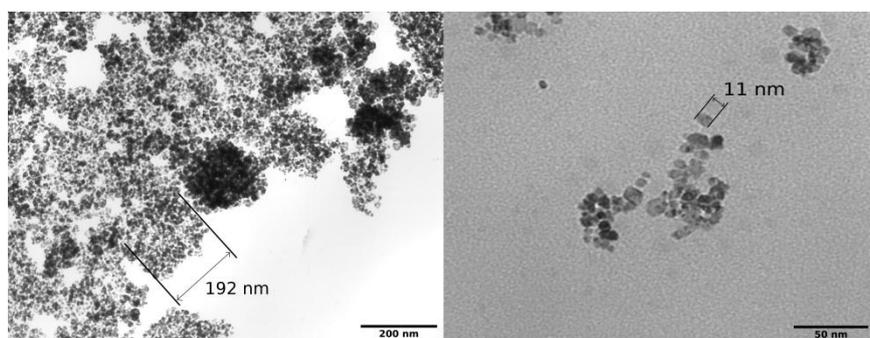

**Fig. 8.** Transmission electron micrographs of FluidMAG-D magnetite nanoparticles. (A,B) nanoparticle solution dried on a coated grid. (A) Aggregated nanoparticles; a putative multicore

nanoparticle is highlighted. (B) Constituent nanoparticle structures.

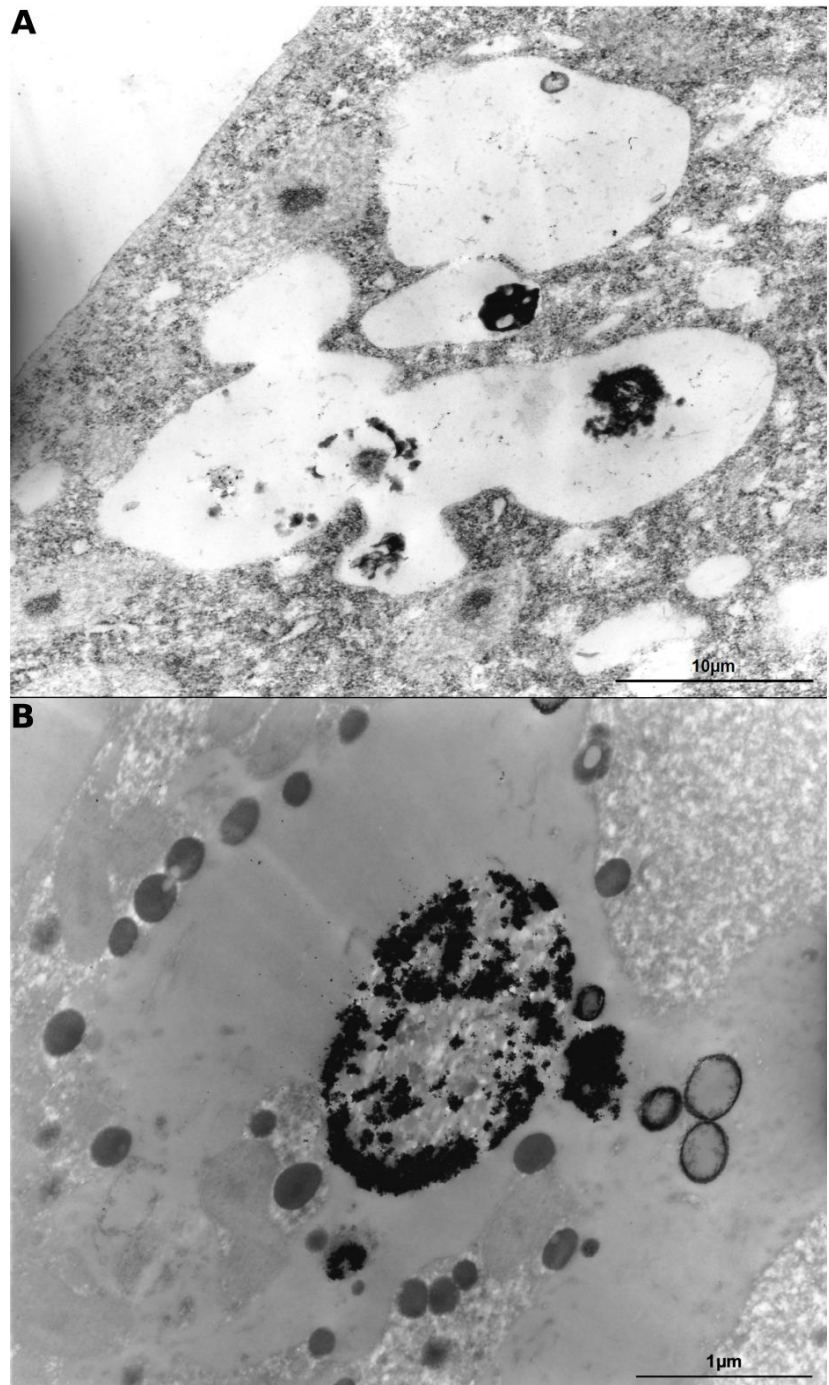

**Fig. 9.** Transmission electron micrographs of magnetite nanoparticles inside the *P. polycephalum* plasmodium. (A) Micrograph showing *in vivo* vacuolised magnetite aggregates in the ectoplasm of plasmodium treated with nanoparticle suspension. (B) *In vivo* iron deposit continuous with the endoplasm.

**Silver-coated Glass Microspheres**

Figure 10 shows dried tracks left in the wake of the moving plasmodium (hereafter referred to as 'empty tubes'); the bright spherical regions interspersing the desiccated tube structures were confirmed to be the silver-coated glass microspheres using EDX (data not shown). Almost all of the beads are visible on the plasmodium; the few which are on the agar are likely to have fallen off as the sample dried.

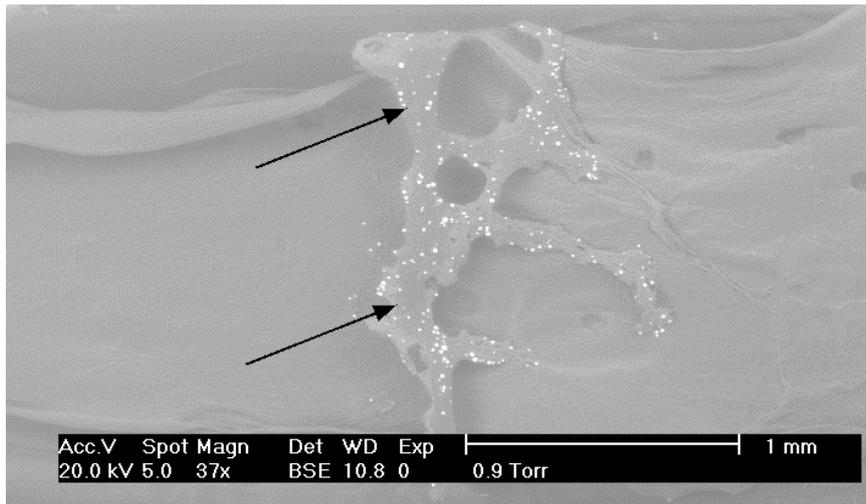

**Fig. 10.** ESEM image of dry empty tubes (arrows) on agar. Silver coated glass spheres are seen as bright circles.

A fully hydrated empty tube is seen in Figure 11. The high BSE signal from the silver coating and glass makes the spheres easy to distinguish from other rounded objects. It is not possible to state whether any slime mould membrane envelopes the spheres (Fig. 12). Occasionally broken tubes were seen where the internal structure was exposed. No spheres were seen in these areas.

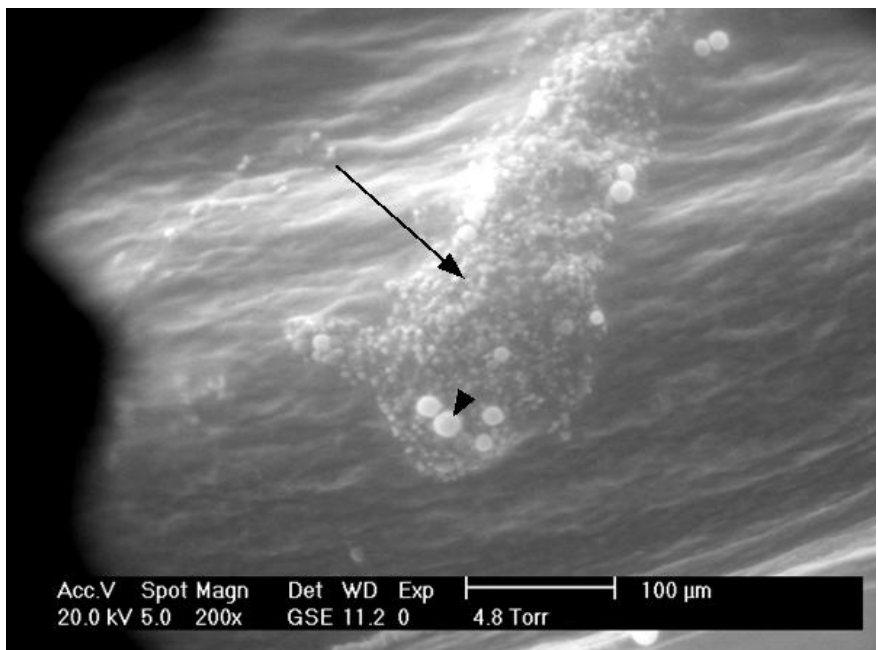

**Fig. 11**. ESEM image showing a fully hydrated empty tube (arrow). Image is derived from an electron secondary detector which has a relatively high BSE component. The enhanced signal makes the spheres (one of which is denoted by an arrowhead) easy to distinguish.

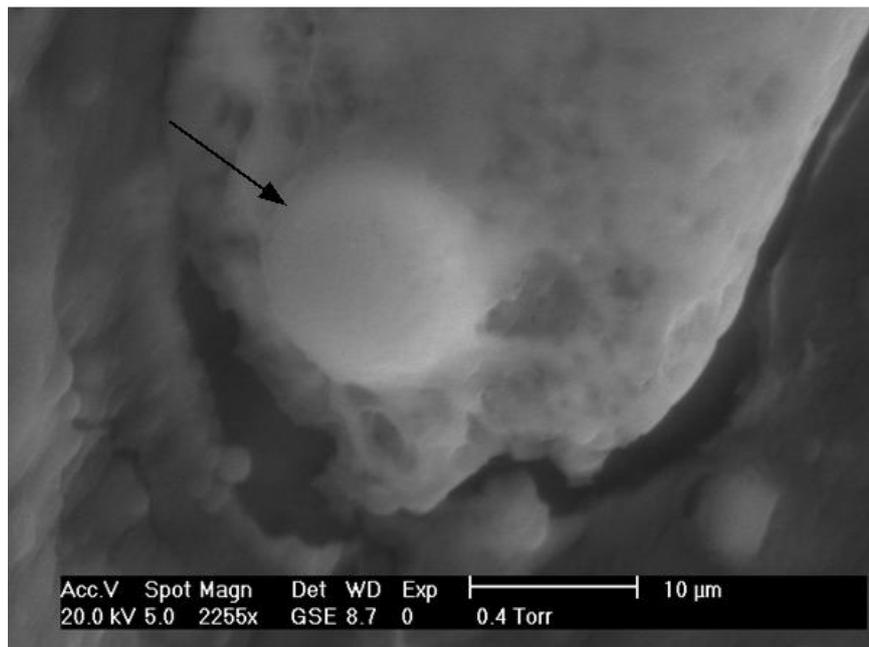

**Fig. 12.** ESEM image showing the tip of a dry empty tube on agar. A silver coated glass bead can be seen in the centre. Dark areas around the tip indicate where shrinkage has occurred.

**DISCUSSION**

Iron deposits seen inside the plasmodium tended to be quite large and probably represent groups of 200nm magnetic particles (Fig. 9). These particles appear to retain some semblance of their native multicore structures, comprised of single domains with a diameter of 5-20nm each (Schaller *et al.*, 2009 and Figs. 8 & 9). This allows us to conclude with confidence that *P. polycephalum* is able to internalise magnetite nanoparticles; their presence in larger quantities in ectoplasmic vesicles suggests internalisation via endocytosis, *P. polycephalum's* mechanism for consuming nutrient sources. The size and structure of plasmodial magnetite deposits are highly discontinuous and not yet fully explained, although the fact of their presence in micrometer-scale quantities suggests some form of assembly: whether this assembly is a consequence of being inside the plasmodium or a spontaneous event is unclear, as are the factors that dictate the size of the resulting structures. These nanoparticles, whilst undeniably biocompatible (i.e. they did not kill the slime moulds which internalised them), did appear to alter the behaviour of the plasmodium slightly.

The silver-coated glass microspheres, however, were not observed as being internalised. This may have been due to the broad spectrum biocidal action of silver. Indeed, previously the repellent effects of silver nanoparticles have been reported (Matveevaa *et al.*, 2006) when using *P. polycephalum* in a chemotactic assay to assess the biological effects of silver nanoparticles. In the aforementioned paper, it was found that *P. polycephalum* actively avoids silver nanoparticles: this is true even when a binary choice between a silver salt (silver nitrate) and silver nanoparticles is set up. In the experimental set up used in this study, *P. polycephalum* does not have the choice of avoiding the silver coated microsopheres, so rather than internalise them (as with magnetite), the plasmodium transports them externally. Without further experiments we cannot be sure of the exact mechanism for this behaviour. It is possible that the plasmodium is simply moving away from the loading zone due to the presence of silver but inadvertently transports particles externally. It is equally possible that the plasmodium is attempting to detoxify the environment by diluting the concentration of silver particles in specific areas of the reaction zone. The silver microspheres may preferentially stick to the outer walls of *P. polycephalum* via a chemically induced reaction to generate excess protective slime layer, and our highest magnification observations would seem to support this hypothesis (Fig. 12). Alternatively, external transport may be observed because the particles are too large to be internalised by vacuolisation mechanisms. In either sense, it was assumed that silver spheres stick to the slime layer of the plasmodium and are dragged by the growing tubes away from the loading domain. The spheres remain in place even when the plasmodium abandoned its' tubes (Figs. 10-12). A higher magnification image (Fig. 12) is difficult to interpret; one cannot rule out the possibility that plasmodial material extends around it.

Githens and Karnovsky (1973) suggested that the optimal size of objects to be internalised by endocytosis is 1 μm; whilst their findings were in a different slime mould species, this is more or less in line with the findings of this study, i.e. 200nm nanoparticles were internalised whereas microspheres >10μm were not. In the same article, Githens and Karnovsky also implicitly demonstrated transcytosis in parallel with endocytosis, i.e. the internalisation of exogenous substances, followed by their excretion; again, this appears to concur with the results obtained in this study.

We propose therefore that *P. polycephalum* utilises up to three ways of relocating nano- and micro-scale objects about the plasmodium: (a) endocytosis, when the plasmodium internalises small and/or biocompatible exogenous objects and distributes them throughout the protoplasm, (b) transcytosis, when objects are transported through the protoplasm and excreted to the external regions of the plasmodium (the ectoplasm and/or peripheral slime layer) some distance away from the loading area, and (c) dragging, when larger and/or toxic objects are stuck to walls of the protoplasmic tubes and propagate together with the growing tubes.

The results of the scoping studies presented here could be further developed according to the following directions. Firstly, the *in situ* nano-assembly of novel devices and computing circuits, i.e. if a slime mould internalises nanoparticles and assembles them into a device with a functionality distinct from that of the native particles. Potential outcomes of such technology could include building intracellular circuitry or the development of metallised artificial skeletons for living cells. More specifically, there is an opportunity for *in situ* assembly of memristors (resistors with memory), as magnetite nanoparticles arranged in arrays have been shown to exhibit a substantial bipolar resistance switching (Kim et al, 2009) which may indicate their memristive nature. The analogous nature of memristive devices to mammalian neurons therefore suggests that this technology, if suitably developed, could be of value to both computing and biomedical discipline.

The second potential application is the development of *in situ* nanosensors. Intake of nanosensors in slime moulds may be controlled via temperature, e.g. in studies on uncoupling endocytosis and exocytosis in thermo-sensitive mutants of slime mould (Labrousse and Satre, 1997). Furthermore, intracellular sensing devices attached to magnetic particles may be manipulated with magnetic tweezers (Shekhar *et al.,* 2010); Figure 13 demonstrates the ease with which such control may be exerted at macro-scale by placing a strong magnet in the vicinity of a plasmodium loaded with magnetic nanoparticles.

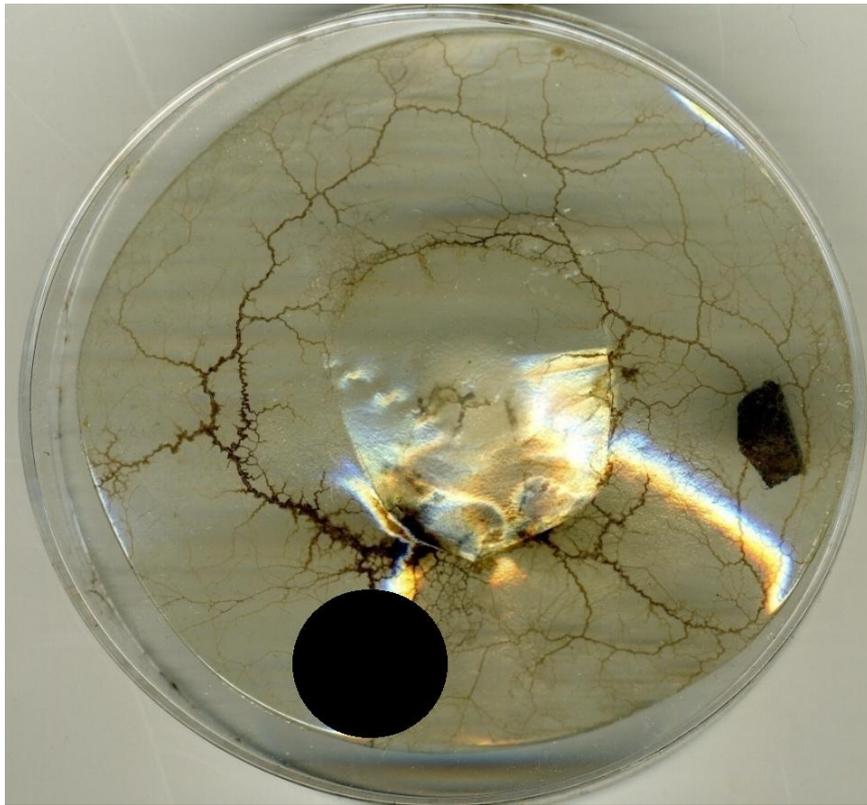

**Fig. 13.** Photograph to demonstrate the control of magnetic nanoparticles in a live protoplasmic network. Position of 25x20mm N52 neodymium magnet is shown by grey disc. Segments of the tubes closest to the magnet exhibit a dense black colour, indicating a high concentration of the internalised magnetic material.

The third potential application is the fabrication of solid-state wires from slime mould's protoplasmic tubes loaded with conductive nanoparticles. In Adamatzky (2010b) we demonstrated that it is possible to program, with tactically positioned sources of nutrients, the propagation of a plasmodium such that the slime mould can deterministically transfer and mix bio-compatible substances. A basic technique for developing slime mould wires is as follows: inoculate slime mould at specific sources, place chemo-attractants in the destination and place chemo-repellents as obstacles in domains to be avoided by plasmodium. Load a plasmodium with a mixture of conductive beads at the source and allow it to develop a protoplasmic tube connecting the source with the destination. The internalised conductive beads are distributed in the protoplasmic tube and dragged on the outer walls of the tube during its growth. When sources of nutrients are exhausted the plasmodium 'abandons' the protoplasmic tubes but the conductive beads remain, e.g. as in Fig. 11: the trace of the conductive beads represents the permanent wire. To link this third potential application with the first, if the plasmodium could be loaded with multiple types of nanoparticle, any potential solid-state device laid down by a propagating plasmodium could be designed with specific properties, e.g. interspersed with micro-scale, plasmodium-assembled resistors, capacitors etc.